# Network Formation in Dairy Gels Revealed with Time-Resolved Very Small-Angle Neutron Scattering

Tianhao Ge[1, 2], Wendeng Li[1, 2], Chun Lei[1, 2], Qing Chen[1, 2], He Cheng[1, 2], François Boué[3], Christopher J. Garvey[4,5*], Guang Wang[1, 2*]

*1. Institute of High Energy Physics, Chinese Academy of Sciences, 19B Yuquan Road, Beijing 100049, China.*

*2. Spallation Neutron Source Science Centre, 1 Zhongziyuan Road, Dongguan 523803, China*

*3. Laboratoire Léon Brillouin, CNRS-UMR12-CEA-Université Paris-Saclay, 91190 Gif-sur-Yvette Cedex, France*

*4. Heinz Maier-Leibnitz Zentrum (MLZ), Technische Universität München, Lichtenbergstraße 1, 85748 Garching, Germany*

*5. Department of Physics, School of Science, RMIT University, Melbourne Vic, Australia*

*Corresponding Authors:

christopher.garvey2@rmit.edu.au

wangguang@ihep.ac.cn

**Abstract**

Two milk gelators, glucono-δ-lactone (GDL), a food grade acidulant, and a commercial chymosin, endopeptidase, have been used to produce milk gels from skimmed milk powder dispersed in $D_2O$. The very small angle neutron scattering, VSANS, instrument at the Chinese Spallation Neutron Source (Dongguan, Guangdong, Peoples Republic of China) has provided access to slit smeared scattering cross-sections for a range of scattering vectors, ($Q = (^{2\pi}/_{\lambda}) \sin(^{\theta}/_{2})$, where λ is the wavelength of the scattered radiation and q is the scattering angle, $5.3 \times 10^{-4} < Q < 0.007$ Å$^{-1}$), and length-scales suitable for understanding the fractal dimension of the network formed by casein micelles (CM) and early stages of CM cluster growth in gels.

The multi-slit collimation system of the time-of-flight VSANS instrument in conjunction with event mode acquisition enables variable temporal resolution of the early stages of gel formation from a sol indicating the growth of clusters. The fractal dimension of cluster growth is consistent with diffusion limited reaction.

## Introduction

Milk is a complex colloidal system in which proteins, minerals, lipids, lactose, and water are organized across multiple length scales. Casein micelles (CMs) are the primary colloidal particles responsible for many of milk's physicochemical and functional properties, including its stability, flow behaviour, and ability to form colloidal scale soft particulate gels (Corredig, et al., 2019; Dalgleish & Corredig, 2012). These hydrated and polydisperse particles consist mainly of αs1-, αs2-, β-, and κ-caseins (phosphoproteins) associated with colloidal calcium phosphate (CCP) (Corredig, et al., 2019; 2020). Scattering studies indicate that bovine casein micelles are approximately spherical, with characteristic dimensions ranging from tens to hundreds of nanometres and structural heterogeneity and containing large amounts of solvent (Dalgleish, et al., 2012; Holt, de Kruif, Tuinier, & Timmins, 2003; Horne 2020; Lenton, Nylander, Teixeira, & Holt, 2015). The colloidal stability of CMs is largely maintained by steric and electrostatic repulsion at the micelle surface. Dairy gelation involves the aggregation of CMs into a 3-dimensional network over time, with reported fractal dimensions (Bauland, Manna, Divoux, & Gibaud, 2024; Breunig, et al., 2026; Ramya, de Campo, Strobl, Boué, & Garvey, 2026; van Heijkamp, et al., 2010) consistent with reaction-limited aggregation mechanisms of aggregation (Teixeira, 1986). Such networks may contain levels of hierarchical structure and connectivity (Bantawa, et al., 2023) where an understanding of network growth is key to understanding the final mechanical properties. Understanding this process and the resulting structures is hampered by a lack of information on the temporal evolution of the 3-dimensional network structure. An experimental description of the network structure's evolution is central to explaining the mechanical properties dairy products formed by the gelation process (Lucey, 2002a).

CMs are commonly described as protein-rich colloidal particles containing nanoclusters of amorphous calcium phosphate. The caseins form a flexible, intrinsically disordered matrix, while phosphorylated regions interact with calcium and phosphate, allowing CCP to be incorporated within the micelle (Holt, Carver, Ecroyd, & Thorn, 2013; Holt, et al., 2003). Small-angle X-ray and neutron scattering studies report a heterogeneous micellar structure containing dispersed calcium phosphate nanoclusters, although the precise organization of these components remains under discussion (Holt, et al., 2013; Lenton, et al., 2015).

The colloidal stability of the CM sol (milk) results from a balance between attractive and repulsive interactions. κ-Casein is particularly important because its hydrophilic C-terminal region extends into the serum and provides steric and electrostatic stabilization, limiting close approach between neighbouring micelles (Holt, et al., 2003; Salvador, Acosta, Zamora, & Castillo, 2022). Interactions between caseins and CCP also contribute to stability, and changes in pH or mineral equilibrium can alter the association between the protein matrix and the mineral phase (de Kruif & May, 1991; Holt, et al., 2013). Milk gelation occurs when the balance favours attractive interactions facilitating micellar aggregation and the formation of a space-spanning network. The destabilisation mechanism influences the aggregation kinetics and resulting network structure across multiple length scales(Li & Zhao, 2019; Salvador, et al., 2022; Yang, et al., 2023).

Enzymatic and acid-induced coagulation are two possible routes to casein gelation. Although both ultimately involve micellar aggregation and network formation, they destabilize the micellar sol stability through different mechanisms (Li, et al., 2019; Salvador, et al., 2022). During enzymatic gelation, chymosin specifically hydrolyses κ-casein at the micelle surface. This cleavage removes the hydrophilic portion while leaving para-κ-casein associated with the micelle, progressively reducing steric stabilization (Adams, et al., 2019; Yang, et al., 2023). Once sufficient κ-casein has been hydrolysed, attractive interactions dominate, allowing micelles to aggregate and form a 3-dimensional network. Chymosin-induced coagulation is commonly described as comprising an initial enzymatic phase followed by a non-enzymatic aggregation phase. The aggregation process depends not only on the extent of κ-casein hydrolysis but also on micelle concentration, ionic conditions, pH, and temperature (Adams, et al., 2019; Horne & Lucey, 2014; Salvador, et al., 2022). Both processes involve a modulation of the interaction potential between CMs in a time dependant fashion: chymosin gelation is related to the enzymic activity, and GDL to the change in bulk pH.

In acid-induced gelation GDL hydrolyses gradually to gluconic acid, producing a decrease in pH (Liu, et al., 2017; Martin, et al., 2009). Acidification changes the ionization state of the caseins and shifts the equilibrium of calcium and phosphate associated with the micelles (Koutina & Skibsted, 2015; Liu, et al., 2017) - these observations imply that there is some buffering capacity in the CM. In particular, decreasing pH promotes dissolution and redistribution of CCP, thereby modifying protein-mineral interactions (de Kruif, et al., 1991; van Heijkamp, et al., 2010). As the pH approaches the isoelectric point of casein, which is approximately 4.6, the net negative charge of caseins decreases, reducing electrostatic repulsion and promoting aggregation through attractive protein-protein interactions (Andoyo, Guyomarc'h, Burel, & Famelart, 2015; Leocmach, Nespoulous, Manneville, & Gibaud, 2015). Micelle aggregation then proceeds until an interconnected acid-induced gel network is formed (Salvador, et al., 2022; van Heijkamp, et al., 2010). These distinct sol destabilisation mechanisms provide a basis for investigating whether chymosin- and GDL-induced gelation produce different structural signatures and aggregation kinetics.

Small angle scattering provides a Fourier perspective on the density correlations within the 3-dimensional structure over the macroscopic volume illuminated by the probing radiation via the absolutely scaled differential scattering cross-section, $d\sigma/d\Omega$, or the likelihood of a photon or neutron being scattered through a particular solid angle normalised to the sample volume (Glatter & Kratky, 1982). In the case of small angle neutron scattering the scattering length density (SLD) fluctuations are defined by the atomic coherent cross-sections in a representative volume. In these soft matter system the SLD is dominated by the scattering power of the $^{1}$H nucleus, and substitution of $^{1}$H for $^2$H enables enhanced contrast/scattering power the ability to highlight particular aspects of structure (Higgins & Benoit, 1994). Using compositional information as a starting point for the analysis, SANS data may be interpreted qualitatively in terms of the length-scales of heterogeneity for samples of well defined composition. Milk is quite a complex sample in terms of chemical composition the dominance of hydrogen facilitates meaningful approximations.

Small-angle neutron scattering (SANS) over a broad range of scattering vectors is well suited to studying CM gels because it non-destructively probes density fluctuations in 3-dimensional structures across the nanometer-to-micron range relevant to individual micelles, aggregates, and developing gel networks (Bouchoux, et al., 2015a, 2015b; Yang, et al., 2023). Extant SANS instrumentation accesses a limited range of scattering vectors determined by the ability to provide definition of the scattering angle, the solid angle of the detector and incident wavelengths. While the distinction is somewhat arbitrary the instrumentation is usually categorised (in order of decreasing scattering vector/increasing length-scales) as: conventional pinhole SANS; very small angle neutron scattering (VSANS) which uses a variety of solutions to overcome the limitations of the pinhole geometry; and ultra small angle neutron scattering (USANS). USANS specifically refers to instrumentation which utilises the Bonse-Hart technique to achieve extremely high angular resolution, and access to the largest structures, for the analysis of the scattered neutron beam at low angles close to the transmitted beam a point at time (de Campo, Garvey, Muzny, Rehm, & Hanley, 2019; Ramya, et al., 2026; Rehm, et al., 2018). SANS and VSANS utilise linear or 2D detectors making them, if measurements are not limited by the scattered intensity, more suitable for kinetic measurements. Depending on the scattering contrast and accessible scattering-vector range, SANS in general can provide information around the internal structure of the CM, the micellar form factor and the structure factor associated with interparticle interactions (Adams, et al., 2019; Bouchoux, et al., 2015b). The VSANS measurements analysed in this study utilised a range of scattering vectors $5.3 \times 10^{-4} < Q < 0.007$ Å$^{-1}$, corresponding primarily to probing micelle-scale, and larger, aggregates.

A major advantage of neutron scattering for milk systems is the strong and adjustable contrast between hydrogen-containing components and the solvent. In whole milk $D_2O/H_2O$ contrast variation can enhance the scattering contributions from fat globules and can help distinguish contributions from protein and mineral components (Lopez-Rubio & Gilbert, 2009). In the present analysis suspensions of skimmed milk powder reconstituted in $D_2O$, we previously defined a two-phase approximation for quantitative analysis of the scattering data in the USANS regime (Ramya, et al., 2026) where two phases are CM and a solvent phase. The CM scattering length density (SLD) is based on literature values while the solvent SLD was defined from a careful consideration of the composition of skimmed powder and the range of scattering vectors.

Previous neutron-scattering studies have demonstrated the potential of time-resolved SANS and USANS for monitoring acid-induced milk gelation and aggregate growth (Yang, et al., 2023). Recent studies of cow, goat, and sheep milk have linked the evolution of CCP-related scattering features with structural changes induced by sample acidity (Adams, et al., 2019; Alexander, Nieh, Ferrer, & Corredig, 2011; Callaghan-Patrachar, Peyronel, Pink, Marangoni, & Adams, 2021). These findings demonstrate that neutron scattering can connect structural changes from the micellar level to the larger-scale aggregates that form the gel network. However in order to access the length-scales associated with aggregation of CMs and formation of the network, which are necessarily larger than the CM length scale ($q{\sim}0.001$ Å$^{-1}$), with the appropriate scattered intensity for temporal resolution of changes in SANS there are significant limitations in sample size. This study uses a multi-slit VSANS approach (Barker, et al., 2022) combined with time-of-flight SANS (Zuo, et al., 2024) to investigate CM aggregation in skimmed-milk/$D_2O$ systems over a range of milk concentrations and during gelation induced by chymosin or GDL. Firstly, non-gelled reconstituted skimmed-milk samples with different concentrations of milk powder are examined to establish how concentration influences the initial micellar structure prior to gelation and interparticle interactions. The modelling of these measurements provides a structural reference, average CM size and size distribution, to precisely describe the t =0 structure for interpreting changes during gel formation. Secondly, time-resolved VSANS is used to follow the structural evolution of casein micelles during chymosin- and GDL-induced gelation. The time-resolved scattering profiles were analysed to compare changes in micellar organisation, interparticle correlations, and aggregate structure during the two gelation processes.

Recently a comparison (Ramya, et al., 2026) between scattering based direct structural interrogation of a fractal network (Chen & Teixeira, 1986) by the Bonse-Hart based ultra small angle neutron scattering and structure inferred from oscillatory rheology interpreted by constitutive models (Muthukumar, 1989) has yielded similar fractal dimensions but approximately an order of magnitude difference in characteristic length- scale of CM aggregates. The larger length-scale inferred from rheology is outside the range which can be probed in such optically dense materials by the small angle scattering in extant instrumentation (Magerl, et al., 2024). In their rationalisation of this result these authors have suggested a second larger length-scale dominates the mechanical properties of the network. While direct access to visualise 3-d gel structure at these length-scales (> 10's microns) is experimentally demanding integrated approaches with experimental, computational and bulk rheology investigations have yielded consistent understanding (Bantawa, et al., 2023). Thus, in this work we aim to utilise the ability of the slit smeared VSANS instrument (encoding of scattering vector with wavelength and time-of-flight together its detector solid angle coverage) to interrogate small scattering vectors corresponding to the network structure with superior temporal resolution and overcome the fundamental limitations of the Bonse-Hart technique in studying the aggregation of CMs. These measurements aim to provide a comparative description of casein-micelle organisation and aggregation during enzymatic and acid-induced gelation, contributing to a better understanding of the structural evolution of dairy protein gels.

## 2. Materials and Methods

### 2.1 Materials

Skimmed milk powder (Groupe Lactalis, Laval, France) was the gift of Thomas

Croguennec (L'Institut Agro Rennes Angers, France). The important compositional information for calculation of the SLD is shown in Table 1. The SLDs and molecular volumes shown in this table were calculated for small molecules using the program MULCH (Whitten, Cai, & Trewhella, 2008) considering appropriate exchangeable H's when dissolved in $D_2O$, or from literature values (CM's). GDL (CAS number: 90-80-2, purity ≥99%) was bought from Aladdin Scientific Corp. The commercial chymosin enzyme CHY-MAX® was supplied by Chr. Hansen, Hoersholm, Denmark. $D_2O$ (CAS number: 7789-20-0, 99.9 atom % D) was supplied by Sigma-Aldrich. All the reagents were used as received without further treatment unless otherwise stated.

**Table 1** Composition of the skimmed milk powder. The listed component densities were used in calculations of solution volume fractions.

| Composition | Mass % | Density (g/cm$^3$) | SLD (×10$^{-6}$, Å$^{-1}$) |
|---|---|---|---|
| Casein | 27.2 | - | 3.08[a] |
| Whey | 6.8 | - | 2.96[b] |
| Lactose | 53 | - | 3.52[c] |
| $H_2O$ | | 1.00 | -0.56 |
| $D_2O$ | | 1.10 | 6.34 |

a. (Lopez-Rubio, et al., 2009; Smith, 2021)

b. Calculated using MULCH and the crystal structure of ß-lactoglobulin (UniProt P02754).

c. Calculated using MULCH

**2.2 Preparation of Skimmed Milk Dispersions**

Appropriate amounts of skimmed milk powder was dissolved in $D_2O$ to make the stock solution, followed by continuous magnetically stirring at 25 °C for at least 2 hours to ensure homogenous dispersion. Prior to use on the beamline the container was sealed with parafilm to avoid atmospheric $D_2O/H_2O$ exchange, contamination or evaporation.

**2.3 Preparation of Chymosin-Induced Gels**

A stock skimmed milk dispersion in $D_2O$ was diluted to prepare the four samples.

CHY-MAX® solution was added at a ratio of 600 μL per 40 g of milk dispersion. The final milk powder concentrations were 0.150, 0.100, 0.050, and 0.025 g/mL. The addition of chymosin defined the nominal start of gelation (t = 0). Each sample was mixed thoroughly with a pipette before being injected into the cell for neutron scattering measurements.

### 2.4 Preparation of GDL-Induced Gels

GDL powder was dissolved in $D_2O$ to prepare a stock solution. A stock skimmed milk dispersion was diluted into four batches, and the calculated volume of GDL solution was added to each batch to obtain a final GDL concentration of 2 wt.%. The final milk powder concentrations were 0.150, 0.100, 0.050, and 0.025 g/mL. The addition of GDL defined the nominal start of acidification (t = 0). After thorough pipette mixing, the sample was injected into the cell for neutron scattering measurements.

### 2.5 VSANS Measurements

Very-small-angle neutron scattering (VSANS) measurements were performed on the multi-slit VSANS instrument at beamline BL14 of the China Spallation Neutron Source (CSNS, Dongguan, China). The instrument operates using a time-of-flight (TOF) technique and incorporates a multi-slit collimation system consisting of 12 precisely aligned components, each containing eight slits, to focus neutron beams onto a high-resolution gas electron multiplier (GEM) detector. The multi-slit components are aligned using a laser-based alignment technique to ensure accurate beam focusing (Luo, et al., 2021). This configuration enables measurements at very low scattering vectors, with an enhanced neutron flux due to the slit geometry (Glatter, et al., 1982), with an instrumental $Q_{min}$ of approximately $2.8 \times 10^{-4}$ Å$^{-1}$ (T. Zuo, et al., 2024). For these measurements a neutron wavelength range of 6.0-10.5 Å was adopted for the Q-range of the measurements.

Immediately after the addition of chymosin or GDL and mixing by pipetting, one portion of each skimmed milk dispersion was transferred into a sample cell for VSANS measurements. The sample cell had a 1 mm path length, with two quartz windows of 2.0 × 3.5 $cm^2$ and held approximately 1 mL of sample. Measurements were conducted at 25 °C. The first VSANS acquisition commenced approximately 3-5 min after the addition of the coagulating agent.

Time-resolved VSANS measurements were conducted to monitor structural changes during chymosin- and GDL-induced gelation. For each gelation run data was collected in a single event-mode file, where for each detector pixel neutrons were encoded with experimental time and time of flight. The single file was could then be post-measurement binned for optimal signal to noise. For chymosin-treated samples, kinetic scattering data were collected continuously for approximately 2-3 h. A further 10 min measurement was made after an interval of approximately 4 h, giving an observation period to approximately 7 h. For GDL-treated samples, scattering data was collected continuously for approximately 6-8 h, followed by a further 10 min measurement after an interval of approximately 6 h, extending the observation period to approximately 14 h. The exposure time per frame was 20 min, and the nominal kinetic time interval between consecutive frames was 0 min. The kinetic scattering profiles were identified by their acquisition frame numbers. The first frame therefore represents the earliest experimentally accessible stage of gelation. After each measurement the sample was inspected to ascertain the formation of a milk gel.

## 2.6 VSANS Data Reduction and Analysis

The VSANS scattering data were reduced to obtain one-dimensional scattering intensity profiles, $I(q)$, on an absolute scale ($cm^{-1}$). The data-reduction methodology followed the principles described for the BL14 VSANS instrument by Zuo, et al., (2024). In this approach, neutron events recorded by the GEM detector are assigned scattering-vector components using their detector positions and time-of-flight information. For the multi-slit configuration, the one-dimensional scattering profiles are obtained with respect to the horizontal scattering-vector component, with integration over the vertical detector direction. The instrument's published absolute-intensity calibration procedure is based on direct-beam measurements and incorporates corrections for sample transmission, sample thickness, and background contributions.

The reduced scattering profiles were analysed using SasView-6.1.3 (Krzywon, et al., 2026). Owing to the multi-slit collimation geometry, the instrumental resolution function is anisotropic, with different broadening characteristics in the horizontal and vertical directions. Previous investigations have demonstrated that this geometry produces substantial slit-smearing effects that can modify both the shape and absolute magnitude of the measured scattered intensity (Han, et al., 2024; Zhu, et al., 2025). To account for instrumental smearing during model fitting, a custom slit-smearing correction was applied to the model calculated curve using parameters provided by the BL14 beamline: a slit length of 0.020 $Å^{-1}$ and a slit width of 0.0002 $Å^{-1}$. The correction was applied to the theoretical scattering profiles during fitting, without explicitly desmearing the experimental data. This treatment represents an approximation of the instrumental resolution rather than a full implementation of the wavelength-dependent, two-dimensional resolution function but avoids any assumptions of the limiting low-q behaviour implicit in the desmearing process (Lake, 1967).

For non-gelled skimmed milk dispersions, the scattering intensities were fitted using a polydisperse sphere form factor combined with a hard-sphere structure factor (Kotlarchyk & Chen, 1983) using SASVIEW. The spherical form factor describes the scattering contribution from individual casein micelles, whereas the hard-sphere structure factor accounts for intermicellar correlations associated with excluded-volume interactions. The CM radii were assumed to follow a lognormal distribution to account for particle-size polydispersity. Within a simplified form-factor-structure-factor representation, the scattering intensity can be expressed as

$$I(Q) = \phi V(\Delta\rho)^2 P(Q) S(Q) + B \quad \textbf{(1)}$$

where $\phi$ is the volume fraction of casein micelles, $V$ is the micellar volume, $\Delta\rho$ is the scattering-length-density contrast between the micelles and the surrounding solvent, $P(Q)$ is the normalized sphere form factor, $S(Q)$ is the interparticle structure factor, and $B$ is the residual q-independent background due to incoherent scattering.

The SLD value of the casein micelles was estimated based on previously reported compositional information, $3.08 \times 10^{-6}$ $Å^{-2}$. Considering the investigated length -scales, components in solvent contributing to the intensity were whey protein, lactose, $D_2O$, and $H_2O$. Thus, the SLD value of the solvent was calculated using

$$\delta_{solvent} = \delta_{whey}\phi_{whey} + \delta_{lactose}\phi_{lactose} + \delta_{D2O}\phi_{D2O} + \delta_{H2O}\phi_{H2O},$$
$$\phi_{whey} + \phi_{lactose} + \phi_{D2O} + \phi_{H2O} = 1 \qquad \textbf{(2)}$$

where $\delta$ and $\phi$ represent SLD and volume fraction of each component. The SLD value of each small molecule was calculated using MULCH and their volume fractions using the density values shown in Table 1. Volume fraction of casein micelles ($\phi_{CM}$) was calculated using

$$\phi_{CM} = c_{CM}(g/mL) \times voluminosity(mL/g) \quad \textbf{(3)}$$

where $c_{CM}$ represents concentration of casein. Voluminosity is 4 mL/g in this work, suggesting that each gram of dry casein occupied 4 mL of the solution. Volume fraction values of each component are listed in Table 2.

**Table 2** The volume fraction of various components of the reconstituted skim milk in $D_2O$.

| components | $\phi$ | | | |
|---|---|---|---|---|
| | 0.15 g/mL | 0.10 g/mL | 0.05 g/mL | 0.025 g/mL |
| Casein micelle | 0.15 | 0.104 | 0.053 | 0.027 |
| Whey | 0.0084 | 0.0054 | 0.0026 | 0.001 |
| Lactose | 0.0635 | 0.0409 | 0.0198 | 0.01 |
| $D_2O$ | 0.9231 | 0.9505 | 0.9761 | 0.989 |
| $H_2O$ | 0.0050 | 0.0032 | 0.0015 | ≈ 0 |

Following our previous USANS work (Ramya, et al., 2026) the VSANS from the gels were fitted with size distribution of CMs from matching data from ungelled reconstituted skimmed milk combined with the structure factor of Teixeira and Chen (1986) for fractal aggregates of spheres again in SASVIEW using the instrument smearing function described above. The important parameters extracted from the modelling are the fractal dimension of the network, $D_f$, and the correlation length, ζ, of the network.

## 3. Results and Discussion

The scattering curves of the each of the ungelled samples and the evolution of the structure are shown in Fig. 1 A -H. While the ungelled samples in all cases do not show a clear Guinier regime, due to the limitations of the Q-range of the measurements on VSANS and the size distribution of the CMs, there is a general trend of the increasing intensity with increasing milk concentration. All the gelling samples exhibit an increase in the low-q intensity with time. On the log-log plot the slope fitted to this region is approximately -1.8. As observed for our previous work (Ramya, et al., 2026) the time-scale for gelation, from the equilibration of the VSANS curves, using chymosin was faster than for acid based GDL gel.

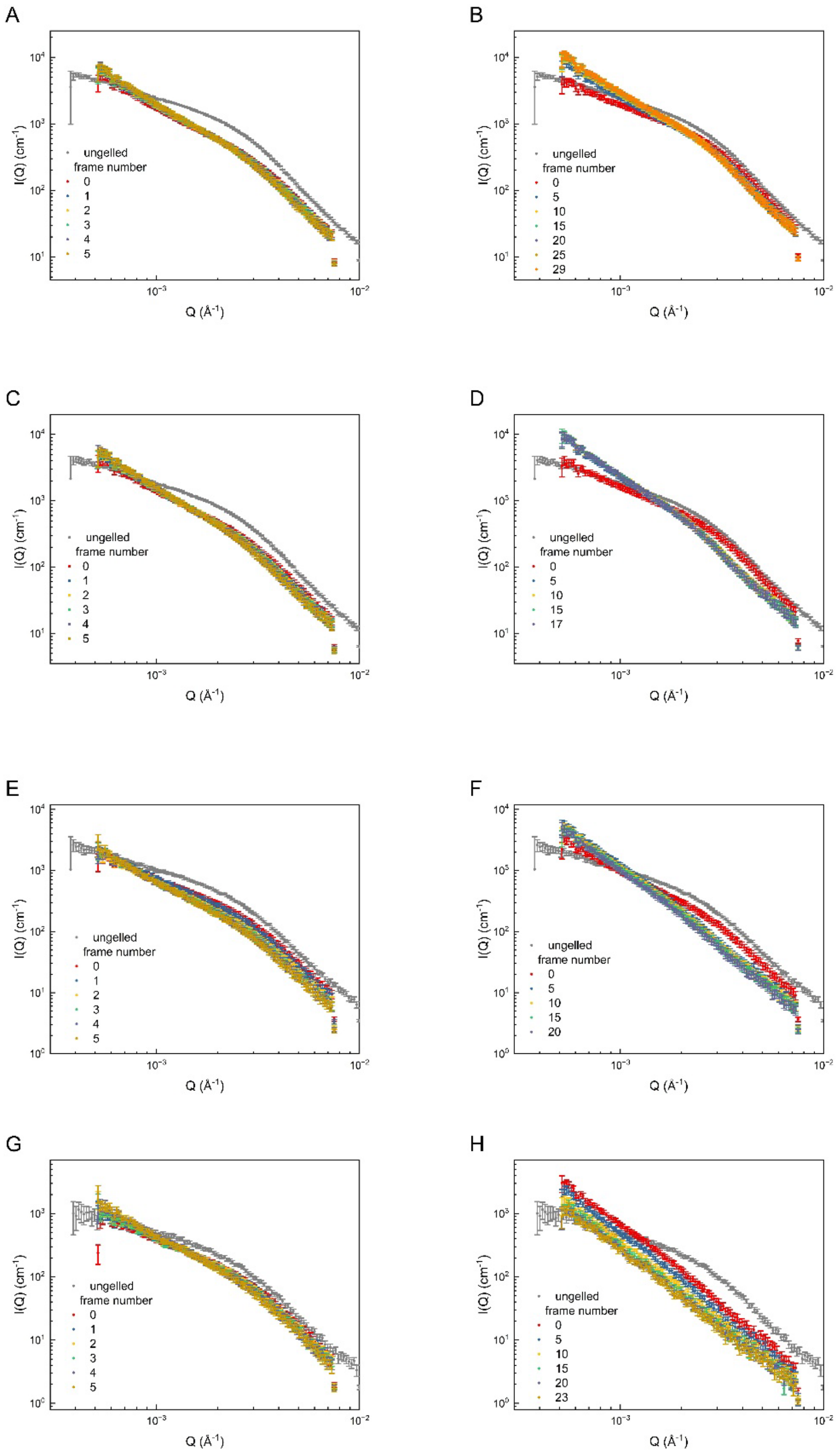

**Fig. 1** VSANS data for chymosin (A 0.15 g/mL, C 0.10 g/mL, E 0.025 g/mL and G 0.05 g/mL) and GDL (B 0.15 g/mL, D 0.10 g/mL, F 0.025 g/mL and H 0.05 g/mL) gels. Each plot has data from the same concentration of ungelled $D_2O$ reconstituted skimmed milk for reference.

### 3.1 Concentration Dependence of the Initial Structure of CMs

The VSANS the non-gelled skimmed milk dispersions was first considered over a series of concentrations (0.025-0.15 g/mL) to establish the structural reference state prior to gelation: size and polydispersity of CMs. Since the VSANS intensities were placed on an absolute scale and the background and sample-container contributions had already been subtracted, the measured intensities were normalized by the corresponding milk concentration, C, to account for the first-order dependence of scattering intensity on the amount of scattering material. The resulting data were represented in Kratky form (Bauland, Manna, Divoux, & Gibaud, 2024) as ${I(Q)Q^2}/{C}$ (Fig. 2).

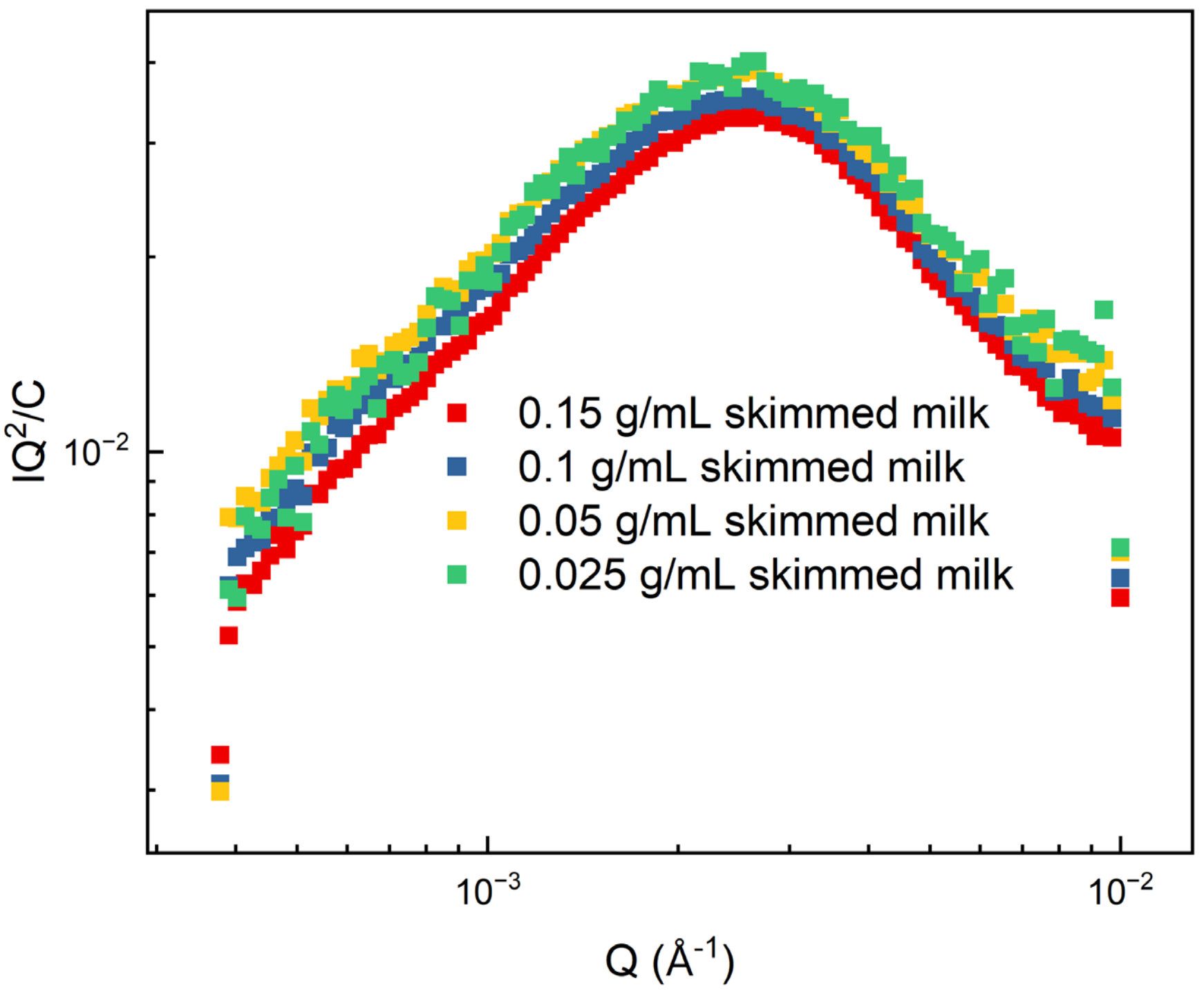

**Fig. 2** Concentration-normalized Kratky representation of the VSANS intensity, ${I(q)Q^2}/{C}$, for non-gelled skimmed milk dispersions prepared in $D_2O$ at concentrations of 0.025, 0.05, 0.10, and 0.15 g/mL.

The concentration-normalized Kratky profiles show a similar shape across all four milk concentrations, characterized by an increase in ${I(Q)Q^2}/{C}$ at low q, followed by a broad maximum at $Q \approx (2.5 - 2.75) \times 10^{-3}$ Å$^{-1}$, and a subsequent decrease toward higher Q. The position of this maximum shows no systematic shift with concentration: $Q_{max}$ values of 0.0027, 0.0025, 0.0027, and 0.0027 Å$^{-1}$ were obtained for milk concentrations of 0.15, 0.10, 0.05 and 0.025 g/mL$^{-1}$, respectively. The preservation of the overall profile and of the characteristic q-position suggests that the dominant structural feature contributing to the scattering remains broadly similar over the investigated concentration range.

Concentration normalization also produces a substantial collapse of the scattering intensities despite the sixfold difference in milk concentration. Nevertheless, the curves do not overlap completely. The maximum values of $I(Q)Q^2/C$ increase from approximately 0.0331 for the 0.15 g/mL sample to 0.0400 for the 0.025 g/mL sample, indicating a modest but systematic concentration dependence after normalization. This residual dependence suggests that the scattering cannot be described solely by a linear increase in the number of identical, noninteracting scatterers.

Within the description of Eqn 1, the similarity of the normalized profiles is consistent with a broadly conserved casein-micelle form factor $P(Q)$, whereas the residual concentration dependence may reflect changes in the interparticle structure factor $S(Q)$ as the micelle concentration increases. In the dilute limit, $S(Q) \rightarrow 1$, the concentration-normalized scattering from structurally equivalent particles would be expected to collapse onto the same curve. Increasing deviations from this behaviour at higher milk concentrations may therefore arise from enhanced intermicellar correlations. Quantitative form- and structure-factor modelling can be used to test this assertion.

This approach is consistent with the scattering framework used by Bauland, et al., (2024) for enzymatic milk gels. In their study, the scattering was represented as $I(Q,t) \propto P(Q)S(Q,t)$, where $t$ is time; the casein-micelle form factor could be maintained during gelation while changes in the structure factor described the development of the casein-micelle network. They further demonstrated that Kratky representation enhances the visualization of changes in scattering associated with structural evolution. In the present non-gelled dispersions, the concentration-normalized Kratky profiles therefore provide a useful reference state for distinguishing the subsequent structural changes associated with chymosin- and GDL-induced gelation.

Fits, refining only parameters to define the size and its distribution, to a log normal polydisperse spherical form factor combined with a hard-sphere structure factor (Sphere × Hard Sphere model - Kotlarchyk, et al., 1983) yielded characteristic radii and radii log normal size distributions shown in Table 3 from the fits shown in Fig. 3. The radii span 43 to 54 nm with a modest decrease at lower concentrations and a slightly decreased polydispersity through the series. These results suggest that the dominant scattering features remained associated with micelle-scale structures, with relatively limited variations in the fitted size distributions.

**Table 3** Parameters obtained from the structural modelling of VSANS for reconstituted skim milk in $D_2O$.

| Concentration (g/mL) | Radius (Å) | Distribution of radii (PD) |
|---|---|---|
| 0.15 | 536±12 | 0.507±0.001 |
| 0.1 | 536±12 | 0.500±0.001 |
| 0.05 | 455±12 | 0.487±0.002 |
| 0.025 | 437±14 | 0.466±0.003 |

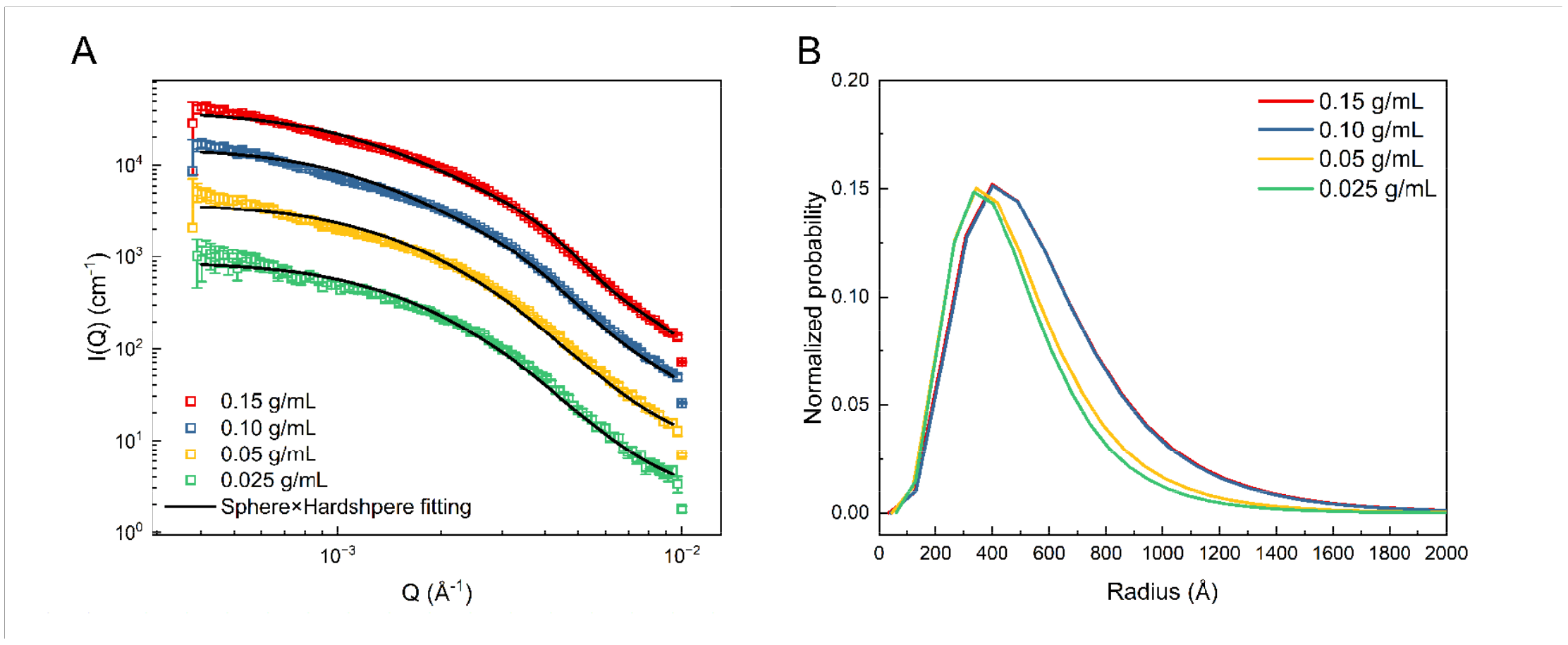

**Fig. 3** (A) Best fits and VSANS data with X-axis shift factors of 2, 4, and 8 applied for 0.05 g/mL, 0.10 g/mL, and 0.15 g/mL skimmed milk dispersions; (B) the corresponding radius lognormal size distribution.

Previous studies have shown that CMs can retain their dimensions over a range of concentrations when the surrounding serum environment is maintained (Nair, Alexander, Dalgleish, & Corredig, 2014). However, dilution with water and the resulting changes in serum composition (e.g. ionic strength) may shift mineral equilibria, micellar hydration, and casein association, potentially affecting micellar structure (Raak, Pedersen, & Corredig, 2023; Wang, Li, Denon, To, & Van der Meeren, 2023). In the present study, all dispersions were prepared by diluting a common $D_2O$-reconstituted milk stock with additional $D_2O$. As shown schematically in Fig. 4, decreasing milk concentration also reduced the concentrations of small molecule milk constituents, potentially modifying the physicochemical environment of the micelles, and in particular the ionic strength.

Furthermore, concentration-dependent intermicellar correlations may influence the structural parameters extracted from the Sphere × Hard Sphere model. At higher micelle volume fractions, stronger interparticle correlations can modify the scattering profiles, while correlations between fitted form-factor and structure-factor parameters may contribute to variations in the extracted radius and PD (De Kruif, 2014; Nair, et al., 2014). Therefore, the observed concentration dependence may reflect a combination of changes in serum conditions and interparticle interactions, rather than changes in micelle dimensions alone.

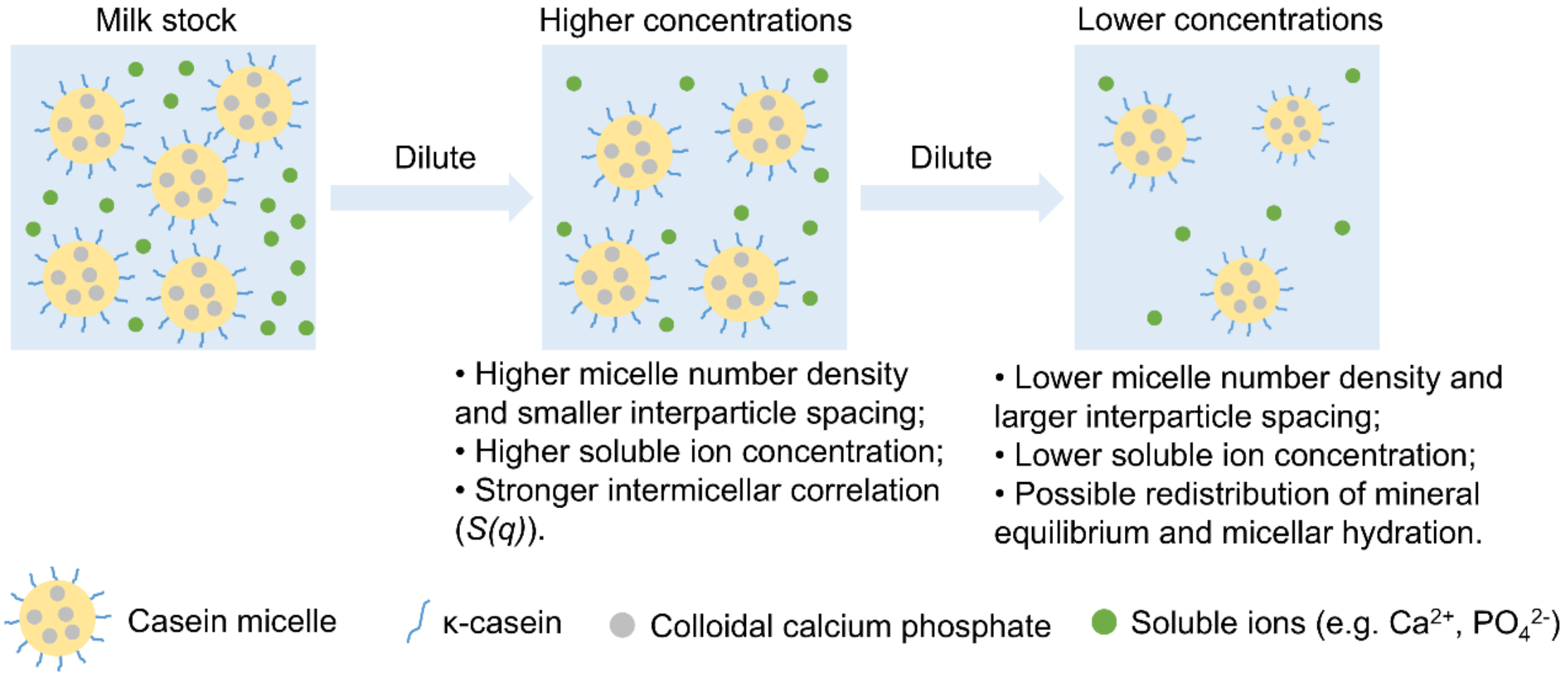

**Fig. 4** Schematic illustration of the proposed concentration-dependent effects of dilution on casein micelle (CM) structure in reconstituted skimmed milk. Changes in micelle number density, interparticle spacing and serum composition may alter intermicellar correlations, mineral equilibria and micellar hydration, potentially influencing the fitted micelle radius and polydispersity.

### 3.2 Structural Evolution during Chymosin-Induced Gelation

The time-dependent structural evolution of skimmed milk during chymosin-induced gelation was investigated using time-resolved VSANS at four milk concentrations (0.025, 0.05, 0.10, and 0.15 g/mL – Fig.1A, 1C, 1E and 1G). The scattered intensities are represented in a fractal-compensated form (Bauland, et al., 2024), $I(Q,t)Q^{1.8}$, to highlight changes in the distribution of scattering intensity across the investigated length scales in Fig. 4. The exponent 1.8 was taken from fitting the linear region, low-Q on the log(I(Q)) versus log (Q) of the fully equilibrated gel structures in Fig. 1. For an unsmeared mass-fractal structure exhibiting $I(Q) \propto Q^{-D_f}$, slit smearing reduced the apparent power-law exponent by approximately 1 on the long-slit limit (Adams, et al., 2019; Gu & Mildner, 2016), yielding $I_{slit}(Q) \propto Q^{-(D_f-1)}$. Therefore, multiplication by $Q^{1.8}$ approximately compensates for the slit-smeared power-law decay expected for $D_f \approx 2.8$, allowing the q-dependant deviations from the fractal scaling regime and changes in the scattering profiles to be more readily identified. The first measurements were initiated approximately 3-5 min after chymosin addition, and the subsequent profiles were identified by their acquisition frame numbers (0-5). Therefore, the earliest measured profiles represent the initial accessible stages of the coagulation process rather than the unperturbed milk dispersions, and show clearly changes from the constituted milk powder.

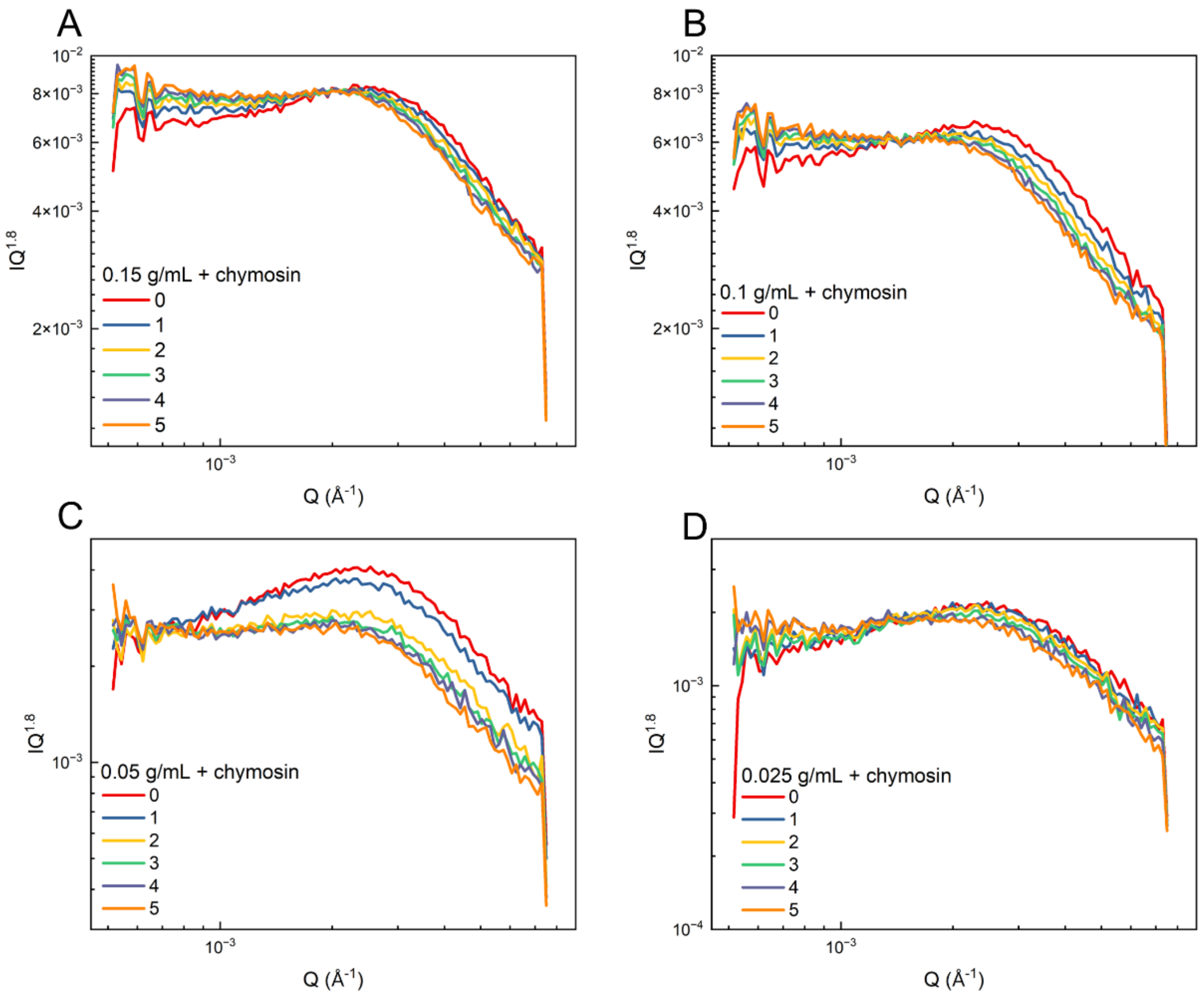

**Fig. 5** Time evolution of the fractal-compensated VSANS profiles, $I(q,t)q^{1.8}$, during chymosin-induced gelation of reconstituted skimmed milk at concentrations of (A) 0.15, (B) 0.10, (C) 0.05, and (D) 0.025 g/mL.

At all four concentrations, the initial fractal-compensated scattering profiles exhibited a broad maximum in the intermediate Q region, followed by a decrease towards higher Q (Fig. 5). During subsequent measurements, the scattering profiles evolved, with the most noticeable changes occurring around and beyond the broad maximum, while increases were also apparent over part of the low-Q region at some concentrations. At the two higher concentrations (0.15 and 0.10 g/mL), the broad maximum remained identifiable throughout the measurement sequence, although its amplitude generally decreased. At 0.05 g/mL, the evolution was more pronounced, with a substantial reduction in the amplitude of the maximum between the initial and later frames. By contrast, the lowest concentration exhibited relatively modest changes across much of the measured Q-range, with considerable overlap between profiles. These observations indicate that the magnitude of the scattering evolution depends on the initial milk concentration, with the most pronounced changes in profile shape observed at 0.05 g/mL.

Chymosin-induced coagulation is initiated by the hydrolysis of κ-casein, which reduces the steric stabilisation of casein micelles and promotes intermicellar association. In the present fractal-compensated representation, the broad maximum and its subsequent evolution reflect changes in the Q-dependent scattering behaviour during coagulation. The relatively flat regions at low-to-intermediate Q are compatible with approximate mass-fractal scaling, although the limited extent of these regions prevents an unambiguous determination of the scaling exponent, $D_f$, from the transformed profiles alone. The gradual reduction in the maximum suggests changes in the relative contributions of different structural correlations, potentially associated with the development and reorganisation of casein micelle aggregates.

A related scattering approach was employed by Bauland, et al. (2024) to investigate the structural evolution of enzymatic milk gels. Their USAXS measurements revealed the development of mass-fractal scattering during gel formation and aging, with the structural evolution described through changes in the interparticle structure factor while maintaining an approximately invariant micellar form factor within the investigated range. In the present study, the evolution of the VSANS profiles is qualitatively compatible with progressive aggregation of casein micelles.

Quantitative fitting of selected kinetic profiles with fractal model of Chen and Teixeira (1986) provides further more quantitative insight into this evolution. The two fitted parameters, $\xi$, the correlation length, and $D_f$, are interpreted as the size of the cluster and the fractal dimension or the compactness of the cluster respectively. The results of this fitting are shown in Fig. 6A. The primary particle radii and lognormal radius polydispersity (PD) were constrained to the corresponding non-gelled values, while $D_f$ and $\xi$ were refined independently for each frame. For the formation of a gel by chymosin from the 0.15 g/mL dispersion, both the $D_f$ and $\xi$ increased slightly. These relatively small variations suggest that the growth of mass-fractal organization remained almost unchanged over the measurement period. This is consistent with the findings of other workers (Bremer, van Vliet, & Walstra, 1989; Callaghan-Patrachar, et al., 2021; Teixeira, 1988; van Heijkamp, et al., 2010). At 0.10 g/mL, $D_f$ increased slightly and there is also a modest increase in ξ. This suggests an increase in mass-fractal compactness without a substantial extension of the characteristic ξ in chymosin gels.

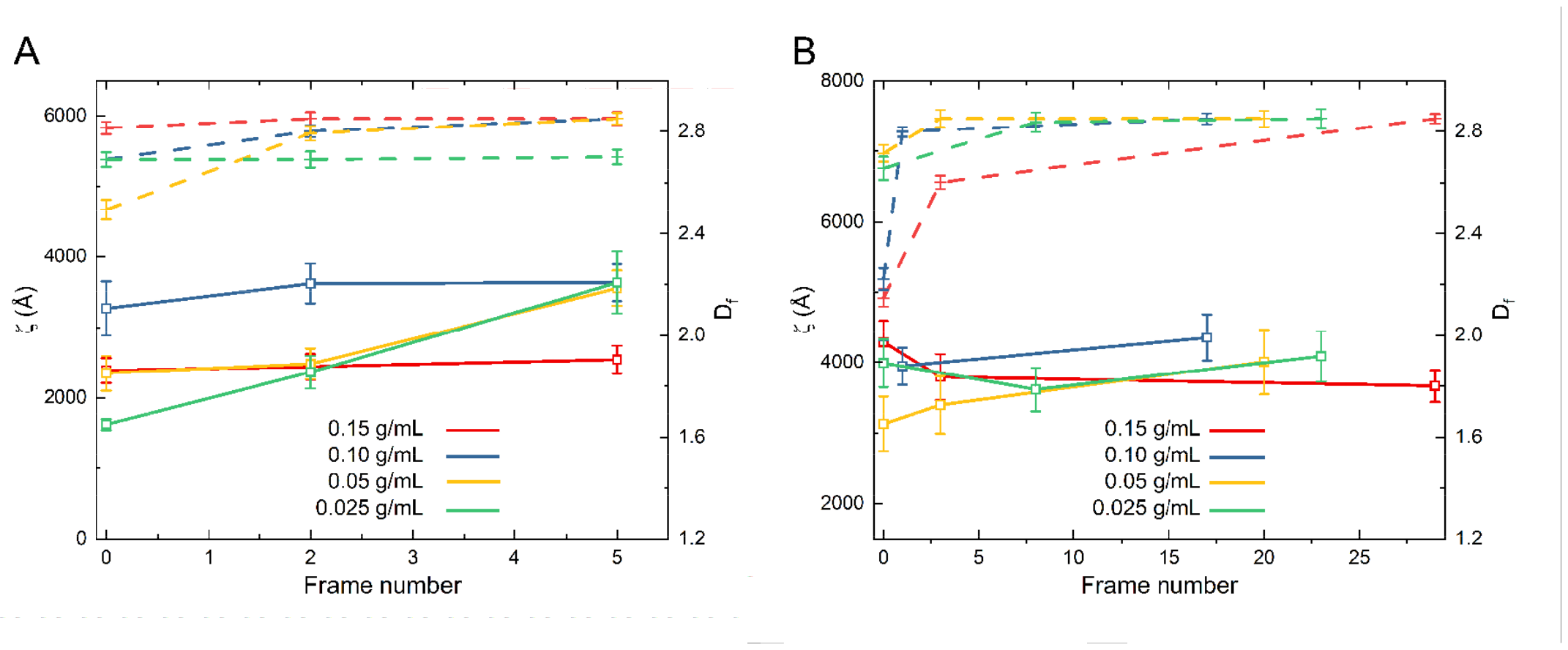

**Fig. 6** Evolution of ζ (solid line and cross) and $D_f$ (dashed line and circular symbol) with frame number obtained from the structural modelling of kinetic VSANS for (A) chymosin- and (B) GDL-induced gels. Colour of lines indicate the skimmed milk powder concentration.

The chymosin-treated dispersion at two lower concentrations, 0.05 and 0.025 g/mL, both exhibited a relatively more marked increase in ξ with modest to no change in $D_f$. In the case of 0.05 g/mL $D_f$ increasing from 2.50 ± 0.04 to 2.85 ± 0.02 and ξ increasing from 2345 ± 246 to 3556 ± 250 Å over the time course of the experiment. This behaviour is consistent with the development of a more compact mass-fractal organization alongside an extension of structural correlations, potentially indicating progressive intermicellar association and structural rearrangement (Mellema, Heesakkers, van Opheusden, & van Vliet, 2000; Mellema, Walstra, van Opheusden, & van Vliet, 2002; Teixeira, 1988; van Heijkamp, et al., 2010). By contrast, at 0.025 g/mL, $D_f$ remained relatively stable at 2.7 ± 0.03, whereas ξ increased from 1624 ± 78 to 3635 ± 435 Å. This suggests that the spatial extent of fractal correlations increased without a corresponding change in mass-fractal compactness. Notably, the three higher concentrations reached similar final $D_f$ values of approximately 2.85, despite differences in their initial values and ξ evolution.

Overall, the fractal-compensated VSANS profiles reveal concentration-dependent structural evolution during chymosin-induced gelation, primarily reflected in changes in the amplitude of the broad maximum. These observations are consistent with progressive reorganisation of intermicellar correlations and suggest that aggregation may proceed through concentration-dependent structural pathways, with changes in fractal compactness and ξ occurring to different extents. It is worth noting that the microscopic interpretation remains model-dependent, particularly where the fitted ξ are not fully constrained by the accessible q-range (Pauw, Pedersen, Tardif, Takata, & Iversen, 2013).

### 3.3 Structural Evolution during GDL-Induced Gelation

The structural evolution of skimmed milk during GDL-induced acidification was investigated using time-resolved VSANS measurements. Four milk concentrations (0.025, 0.05, 0.10, and 0.15 g/mL) were examined to evaluate the influence of milk concentration on acidification kinetics and the associated structural reorganization of casein micelles. The time-dependent scattering profiles were represented in fractal-compensated form $I(Q,t)Q^{1.8}$ (Fig. 7) as introduced in Section 3.2.

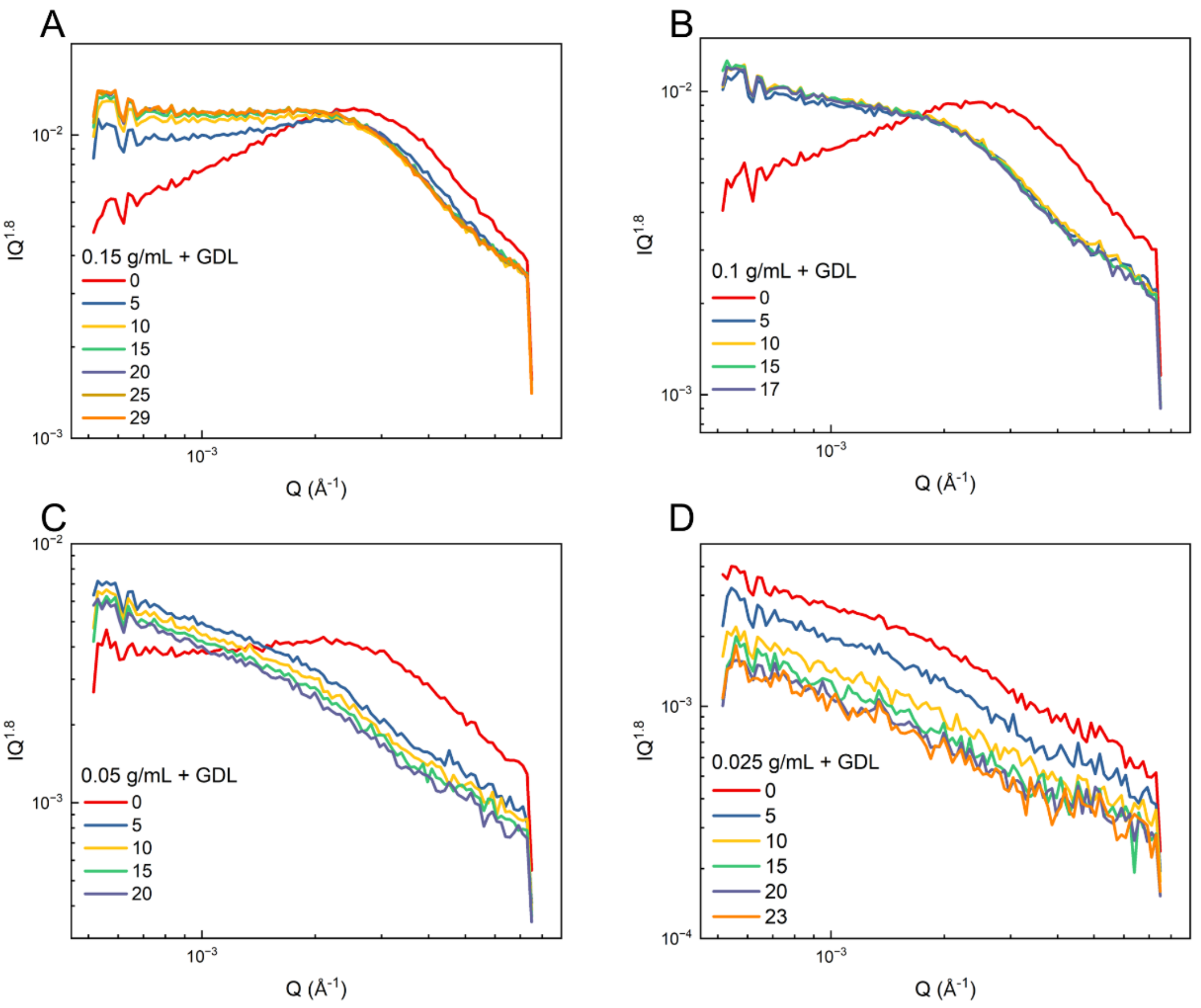

**Fig. 7** Time evolution of the fractal-compensated VSANS profiles, $I(Q,t)q^{1.8}$, during GDL-induced gelation of reconstituted skimmed milk at concentrations of (A) 0.15, (B) 0.10, (C) 0.05, and (D) 0.025 g/mL.

The time-resolved fractal-compensated VSANS profiles revealed pronounced structural evolution during GDL-induced gelation (Fig. 7). At 0.15 g/mL, the initial profile exhibited a broad maximum in the intermediate Q region. During subsequent measurements, this maximum became less distinct, producing a flatter low-to-intermediate Q profile, while the relative contribution from low-Q scattering increased. The later profiles showed comparatively small differences, suggesting that the most substantial changes detectable within the investigated q-range occurred between the initial and subsequent measurement frames.

At 0.10 g/mL, a similar but more pronounced transformation was observed. The initial profile displayed a distinct broad maximum, whereas subsequent profiles exhibited a marked reduction in its amplitude and a relatively enhanced low-Q contribution. The later profiles largely overlapped, indicating comparatively limited further evolution within the accessible observation period. At 0.05 g/mL, the broad maximum also became substantially less prominent after the initial measurement, accompanied by a relative redistribution of scattering intensity towards lower Q. However, the later profiles continued to exhibit a smaller and generally systematic decrease in the transformed scattering intensity across much of the measured range, suggesting ongoing structural evolution. At the lowest concentration, the profiles showed a less clearly defined intermediate Q maximum and a progressive reduction in transformed scattering intensity across much of the investigated Q-range. Unlike the more concentrated samples, differences remained more apparent across several subsequent time frames, although the latest profiles overlapped considerably, indicating a more gradual evolution of the measured scattering structure.

These concentration-dependent changes can be interpreted in the context of acid-induced destabilisation of casein micelles. Progressive acidification modifies the ionisation state of casein proteins, alters electrostatic interactions and shifts the equilibrium of colloidal calcium phosphate. These processes may affect both the internal organisation of casein micelles and their interactions with neighbouring particles. Consequently, the evolution of the scattering profiles may involve coupled changes in the micellar form factor, $P(Q,t)$, and the interparticle structure factor, $S(Q,t)$. It is also worthwhile to note that in each sample, although the concentration of GDL remains the same, the buffering power of the added skimmed milk powder is increasing, and acting against the pH changes induced by GDL hydrolysis.

The relative enhancement of low-q scattering at the higher concentrations is compatible with the development of larger-scale intermicellar correlations during acidification. However, this interpretation is not unique, as changes in micellar structure, scattering contrast and interparticle organisation may contribute simultaneously to the measured intensity. The reduction in the broad maximum reflects changes in the underlying Q-dependent scattering behaviour, but does not independently establish the formation or growth of mass-fractal aggregates. In particular, the progressive decrease in scattering intensity at 0.025 g/mL may reflect a combination of structural reorganisation and contrast-related changes associated with acidification.

The fitted mass-fractal parameters further reveal the concentration-dependent structural evolution during GDL-induced gelation (Fig. 6B). At 0.15 g/mL, $D_f$ increased from 2.15 ± 0.03 at frame 0 to 2.60 ± 0.03 at frame 3 and 2.85 ± 0.02 at frame 29, while $\xi$ decreased slightly from 4289 ± 305 Å to 3792 ± 325 Å and then to 3664 ± 223 Å. This evolution suggests progressive development of a more compact mass-fractal organization without a corresponding increase in the characteristic extent of structural correlations. At 0.10 g/mL, $D_f$ increased from 2.22 ± 0.04 to 2.85 ± 0.02, whereas $\xi$ increased modestly from 3984 ± 330 to 4353 ± 330 Å. This indicates that the structural evolution at higher milk concentrations was primarily reflected in increasing mass-fractal compactness rather than substantial changes on $\xi$.

At 0.05 g/mL, the fitted $D_f$ increased more moderately from 2.71 ± 0.03 to 2.85 ± 0.03, while $\xi$ increased from 3133 ± 384 to 4006 ± 458 Å. This behaviour is consistent with a modest increase in mass-fractal compactness along with an extension of structural correlations. At 0.025 g/mL, $D_f$ increased from 2.66 ± 0.05 to 2.85 ± 0.04, while $\xi$ remained approximately stable, changing from 3990 ± 342 to 4087 ± 361 Å.

Overall, the concentration-dependent evolution of $D_f$ and $\xi$ suggests that changes in mass-fractal compactness and the spatial extent of structural correlations are not necessarily coupled during the acid-induced gelation. Compared with the limited changes at lower concentrations, the more pronounced increase in $D_f$ at higher concentrations may reflect differences in the progression of micellar destabilization and subsequent structural rearrangement.

Overall, the fractal-compensated VSANS profiles reveal substantial concentration-dependent structural evolution during GDL-induced gelation, characterised by changes in the broad maximum and the distribution of scattering intensity across the investigated q-range. The higher concentrations exhibited pronounced changes between the initial and subsequent measurements, followed by comparatively limited evolution, whereas the lowest concentration showed more gradual changes throughout the measurement sequence. These observations suggest that acid-induced micellar destabilisation, aggregation and structural reorganisation may follow concentration-dependent pathways, involving changes in intermicellar correlations.

### 3.4 Comparison of Chymosin- and GDL-Induced Gelation Pathways

Time-resolved VSANS reveals distinct structural evolution during chymosin- and GDL-induced gelation, reflecting differences in their underlying destabilisation mechanisms, as schematically illustrated in Fig. 8. Chymosin-induced κ-casein hydrolysis primarily reduces steric stabilisation and promotes intermicellar association (Choi, Horne, & Lucey, 2007; Glantz, et al., 2010; Lucey, 2002b; Salvador, et al., 2022), whereas GDL-induced acidification additionally modifies protein charge, colloidal calcium phosphate equilibria and interparticle interactions (Li, et al., 2019; Moitzi, Menzel, Schurtenberger, & Stradner, 2011; Nicolai & Chassenieux, 2021). These differences are reflected in the fractal-compensated scattering profiles, where chymosin treatment generally preserves a recognisable broad maximum, while GDL treatment produces more pronounced changes in its shape and relative scattering contributions. The latter behaviour is consistent with coupled changes in micellar organisation and intermicellar correlations, although their individual contributions cannot be uniquely resolved from the measured profiles.

Milk concentration further influences these structural pathways by modifying interparticle separation, encounter frequency and the physicochemical environment of the micelles. At higher concentrations, closer micellar proximity may facilitate interparticle association, whereas dilution alters serum composition and may affect micellar stability and structural rearrangement. Previous studies have demonstrated concentration-dependent changes in rennet-induced gel properties and structural reorganisation, as well as coupled micellar restructuring and aggregation during acidification (Gaygadzhiev, Corredig, & Alexander, 2009; Mellema, et al., 2002; Moitzi, et al., 2011). These effects may contribute to the different structural responses observed across the concentration series.

The fitted mass-fractal parameters provide further insight into these differences (Fig. 6). During chymosin treatment, changes in $D_f$ were relatively modest at higher milk concentrations but more pronounced at 0.05 g/mL. In contrast, GDL treatment produced substantial increases in $D_f$ at 0.15 and 0.10 g/mL, with comparatively smaller changes at lower concentrations. The $\xi$ trajectories also differed. At 0.05 g/mL, both treatments exhibited increasing $D_f$ and $\xi$, consistent with changes in mass-fractal compactness accompanied by an extension of structural correlations. At 0.025 g/mL, chymosin treatment produced a marked increase in $\xi$ with relatively stable $D_f$, while GDL treatment showed increasing $D_f$ with little change in $\xi$. These contrasting trends suggest that changes in mass-fractal compactness and the spatial extent of structural correlations depend on both milk concentration and the destabilisation mechanism.

Despite these differences, both treatments yielded similar fitted $D_f$ values of approximately 2.85 at late stage, indicating comparable mass-fractal scaling within the investigated Q-range. Overall, the results suggest that concentration-dependent changes in interparticle interactions and physicochemical conditions influence the balance between micellar destabilisation, structural rearrangement and aggregation, leading to distinct structural evolution pathways during enzymatic and acid-induced gelation.

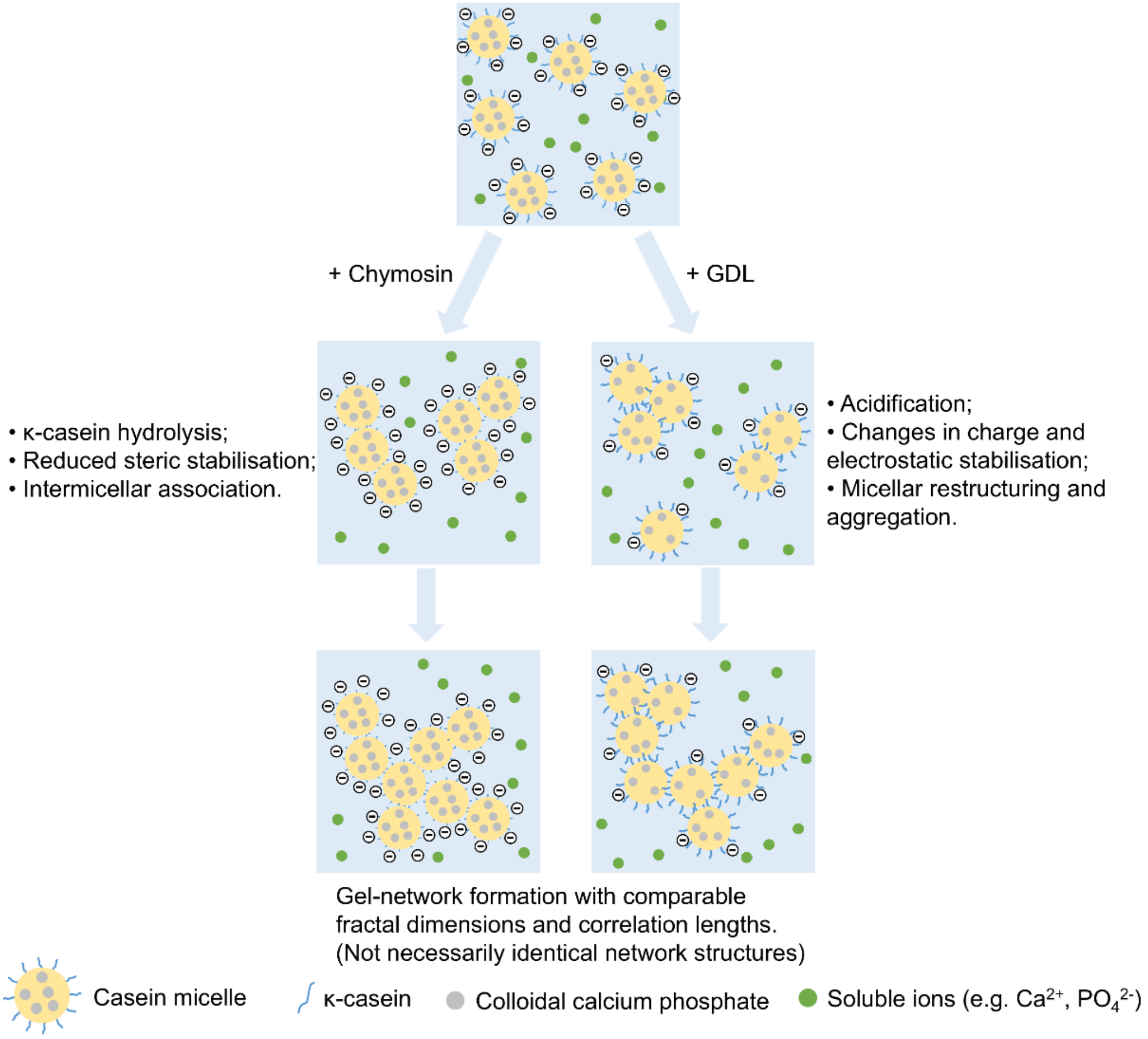

**Fig. 8** Schematic illustration of the proposed structural evolution during chymosin- and GDL-induced milk gelation. Chymosin-mediated κ-casein hydrolysis reduces steric stabilisation, whereas GDL-induced acidification alters casein charge, mineral equilibria and micellar interactions. Both pathways promote intermicellar aggregation and can yield networks with comparable fractal dimensions and correlation lengths.

## 4. Conclusions

This study investigated the concentration-dependent structure of CMs and their evolution during chymosin- and GDL-induced gelation using time-resolved VSANS. For non-gelled skimmed milk dispersions, the fitted micelle radii ranged from approximately 43 to 54 nm, with broadly comparable size distributions across the investigated concentrations. The modest variations in fitted radii can reflect the combined influence of serum conditions and intermicellar correlations instead of changes in micelle dimensions alone.

Both treatments produced concentration-dependent structural evolution through distinct destabilisation mechanisms. The fitted mass-fractal parameters revealed that structural compactness and the characteristic extent of spatial correlations evolved differently depending on milk concentration and the destabilisation mechanism, suggesting that intermicellar association and structural rearrangement may contribute to different extents during gelation. Despite these differences, both treatments yielded similar $D_f$ of around 2.85 at late stage, suggesting comparable mass-fractal scaling within the investigated length scales. However, this similarity does not necessarily imply the same aggregate morphologies or gel-network connectivity/mechanical properties.

Overall, these findings highlight the interplay between destabilization mechanisms, milk concentration, and structural evolution during casein-micelle gelation. Time-resolved VSANS provides valuable insight into these processes across multiple length scales, while the detailed structural interpretation remains subject to the constraints of the scattering model and experimental Q-range. The capabilities of the VSANS instrument, low-Q balanced with an accessible Q-range for kinetic insight, provide not only a unique capability for these studies, but a benchmark for further instrument development at large scale facilities to further enable fundamental insight in an important food production system, the dairy and more generally food gels of different origins.

## Acknowledgements

We thank the staff members of the Very-Small-Angle Neutron Scattering (VSANS) instrument (https://cstr.cn/31113.02.CSNS.VSANS) at the China Spallation Neutron Source (CSNS, https://cstr.cn/31113.02.CSNS) for providing technical support and assistance in data collection and analysis. Garvey would acknowledge the support of the CSNS for travel funding to China during this project and Professor Thomas Croguennec of L'Institut Agro (Rennes Angers, France) for useful discussions.